\documentclass[superscriptaddress,twocolumn,prl,showpacs]{revtex4}

\usepackage{amsmath,amssymb}
\usepackage{graphics}
\usepackage{epsfig}
\usepackage{color}

\begin{document}

\title{Universal Four-Boson States in Ultracold Molecular Gases: Resonant Effects in Dimer-Dimer Collisions}


\author{J. P. D'Incao}
\affiliation{Department of Physics and JILA, University of Colorado, Boulder, CO 80309-0440, USA}
\affiliation{Institut f\"ur Quantenoptik und Quanteninformation, \"Osterreichische Akademie der Wissenschaften, 6020 Innsbruck, Austria}
\author{J. von Stecher}
\affiliation{Department of Physics and JILA, University of Colorado, Boulder, CO 80309-0440, USA}
\author{Chris H. Greene}
\affiliation{Department of Physics and JILA, University of Colorado, Boulder, CO 80309-0440, USA}

\begin{abstract}
We study the manifestations of universal four-body physics in ultracold dimer-dimer collisions. 
 We show that resonant features associated with three-body Efimov physics and dimer-dimer scattering lengths are universally related. The emergence of universal four-boson states allows for the
tunability of the dimer-dimer interaction, thus enabling the future study of ultracold molecular gases with 
both attractive and repulsive interactions. Moreover, our study of the interconversion between dimers and 
Efimov trimers shows that $B_{2}+B_{2}\rightarrow B_{3}+B$ rearrangement reactions can provide an 
efficient trimer formation mechanism. Our analysis of the temperature dependence of this reaction provides 
an interpretation of the available experimental data and sheds light on the possible experimental realization 
of rearrangement processes in ultracold gases.
\end{abstract}

\maketitle 

In both nuclear and atomic physics, the simplest examples of quantum halo states \cite{Halos,HalosGases} 
are weakly bound dimers, with large radii extending well into classically forbidden regions. A remarkable
consequence of their large size is that quantum-halo dimers obey universal scaling laws, i.e., many of their 
properties are independent of the details of their short-range interaction, typically characterized by a 
short length scale $r_{0}$. The notion of a halo state extends in a nontrivial way to more complex 
few-body systems, of which three-body Efimov states \cite{Efimov} are the most prominent example. 
Recently, the notion of universality has been extended to four-boson systems \cite{4bos1,4bos2,Legacy} and 
a new universal picture has emerged. For {\em each} Efimov trimer 
precisely two four-boson states have recently been shown to exist, whose energies are universally related to the Efimov trimer energy \cite{4bos1,Legacy}. 
Thus, the four-body system inherits many of the characteristics of three-body Efimov physics, as the geometric scaling 
of energies and length scales \cite{braaten2006ufb}. More recently, a first experimental evidence of such universal four-body states has been found in an ultracold gas of Cs atoms \cite{Innsbruck}, through the observation of resonant losses due to four-body recombination satisfying the universal predictions of Ref.~\cite{Legacy}. These findings, along with the rapid experimental 
advances in controlling few-body correlations in ultracold quantum gases \cite{Innsbruck,ADRudi,CSdimers,Others}, magnify the importance of four-body scattering processes. They offer a path for the observation 
of universal physics and can potentially enrich the range of experimentally accessible phenomena.

In this Letter, we explore dimer-dimer scattering processes and analyze the consequences of the recently-discovered universal 
properties of four-boson systems \cite{4bos1,Legacy}. Our interest concentrates on such processes near a Feshbach resonance \cite{HalosGases} 
where, through application of an external magnetic field, the $s$-wave two-body scattering length $a$ can be tuned 
from $-\infty$ to $+\infty$. Much of the recent progress in understanding 
dimer-dimer correlations in ultracold physics has been devoted to dimers formed from fermionic atoms \cite{FFdimers}. 
For bosons, the relatively simple fermionic few-body physics is replaced by a much more complex structure involving
multiple few-body halo states that can strongly affect dimer-dimer correlations and the collisional behavior of 
entire gas. Here we advocate that ultracold molecular gases 
are perhaps the best candidates for exploring universal four-body physics at its full complexity. In fact, the search has already begun. In 
Ref.~\cite{CSdimers}, Ferlaino {\em et al.} formed an ultracold sample of Cs$_{2}$ dimers and found by 
varying $a$ that the loss coefficient exhibits a pronounced minimum which allows for longer lifetimes. The temperature 
dependence of the loss rate was also found to display an intriguing behavior, deviating from the expected Wigner threshold law. 

In the present exploration of four-boson universal physics we find that, in contrast to the fermionic case where dimers interact repulsively 
\cite{FFdimers}, bosonic dimers display resonance effects caused by four-body states \cite{4bos1,Legacy}. Near such 
resonances, the dimer-dimer scattering length $a_{dd}$ can vary from $-\infty$ to $+\infty$, thereby resulting effectively in  {\em either} attractive or repulsive interactions. 
Our studies of dimer-trimer conversion, similar in spirit with the one in Ref.~\cite{FXdimers}, further indicate that a molecular gas can be used to form one of Efimov trimers. 
We find that when the atom-trimer collision threshold becomes nearly degenerate with the dimer-dimer threshold, most of the
molecular losses are due to $B_{2}+B_{2}\rightarrow B_{3}+B$ rearrangement scattering. Because of the vanishingly small 
kinetic energy released in such process, it allows trimers to remain trapped. Here we show that the temperature dependence of this 
rearrangement reaction rate can partially explain some recent experimental observations by Ref.~\cite{CSdimers}, which indicate the existence of a trimer
state just above the dimer-dimer collision threshold.

We study dimer-dimer processes using the adiabatic hyperspherical representation for the four-body problem \cite{CGHS}, 
which offers a simple and conceptually clear description of the bound and scattering properties. Here, the Schr{\"o}dinger equation
reduces to a simple system of ordinary differential equations given by, 
\begin{eqnarray}
\left[-\frac{\hbar^2}{2\mu}\frac{d^2}{dR^2}+W_{\nu}\right]F_{\nu}+\sum_{\nu'\neq\nu} W_{\nu\nu'}F_{\nu'}=E F_\nu,
\label{radeq}
\end{eqnarray}
where the hyperradius $R$ describes the overall size of the system, and $\nu$ is a collective index that represents all quantum numbers
necessary to label each channel. In Eq.~(\ref{radeq}), $\mu$ is the four-body reduced mass, $E$ is the total energy, and $F_{\nu}$ is the hyperradial 
wave function. $W_{\nu\nu'}(R)$ are the nonadiabatic couplings that drive inelastic transitions between
channels and $W_{\nu}(R)$ are effective potentials that support bound and quasi-bound
states dictating many of the scattering properties of the system.

\begin{figure}[htbp]
\includegraphics[width=3.2in,angle=0,clip=true]{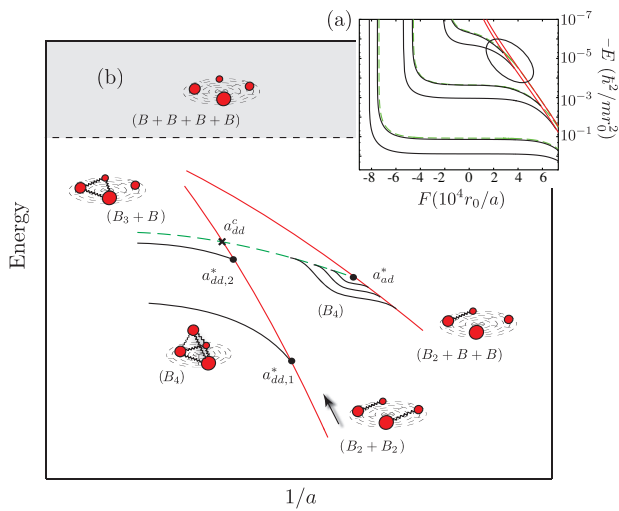}
\caption{(color online). 
(a) Spectrum of the four-boson system from our numerical calculations \cite{Legacy} and (b) a schematic representation of 
the region important for dimer-dimer collisions  (see text).
The four-boson states are represented by black solid lines, while the 
collision thresholds are represented by solid red lines for the atom-atom-dimer and dimer-dimer collisions, and by green dashed lines
for atom-trimer collisions.}
\label{Fig1}
\end{figure}

Figure~\ref{Fig1}(a) shows the spectrum for the two-, three- and four-boson systems from our numerical study \cite{Legacy},
and Fig.~\ref{Fig1}(b) shows a schematic representation of the region marked by a circle in Fig.~\ref{Fig1}(a). [Note that 
we use the function $F(x) \equiv \mbox{sgn}(x)\ln(1+|x|)$ to visualize the whole energy landscape.]
The black solid lines are the energies of the four-boson states, 
the red-solid lines represent the dimer-dimer (lower) and dimer-atom-atom (upper)
collision thresholds, and the green-dashed lines are the Efimov trimer energies, or more precisely, the atom-trimer collision thresholds. 
The structure shown in Fig.~\ref{Fig1}(b)  repeats every time $a$ increases by the Efimov geometric factor $e^{\pi/s_{0}}\approx22.7$, reflecting
the pervasive influence of Efimov physics on the four-boson properties.

Figure~\ref{Fig1}(b) illustrates the resonant effects possible in dimer-dimer collisions ($a>0$).
In general, one expects pronounced features in ultracold scattering observables whenever a bound state crosses, or emerges from, 
the collision threshold, or when an additional decay channel becomes energetically available.
In Fig.~\ref{Fig1}(b), for instance, as $a$ increases and approach $a_{ad}^*$, where an Efimov trimer is created and causes a divergence in the atom-dimer scattering length $a_{ad}$, 
an infinite number of four-boson states \cite{braaten2006ufb} emerge potentially causing resonant effects in 
ultracold atom-molecule gas mixtures \cite{ADRudi}.
For a pure molecular gas, however, it is primarily collisions of the type $B_{2}+B_{2}$ that provide the pathway to detailed study of four-body
physics. As Fig.~\ref{Fig1}(b) shows, there exist precisely two four-boson states emerging from the $B_{2}+B_{2}$ threshold 
at two specific values of $a$, which we denote as $ a_{dd,i}^{*}$ (where $i=1$ and $2$), causing $a_{dd}$ to diverge at values \cite{Jeremy}. In the vicinity of $a_{dd,i}^*$, therefore, the 
dimer-dimer interaction can be tuned from attractive, $a_{dd}<0$, to repulsive, $a_{dd}>0$, irrespective of 
the effectively repulsive character ($a>0$) of the interatomic interactions. At $a=a_{dd}^c$, however, a different process takes
place. Here, the $B_{2}+B_{2}$ and $B_{3}+B$ channels become degenerate [see Fig.~\ref{Fig1}(b)] and one can expect 
an enhancement of the $B_{2}+B_{2}\rightarrow B_{3}+B$ rearrangement reaction  rate over other dimer loss processes. 

\begin{figure*}[htbp]
\includegraphics[width=6.2in,angle=0,clip=true]{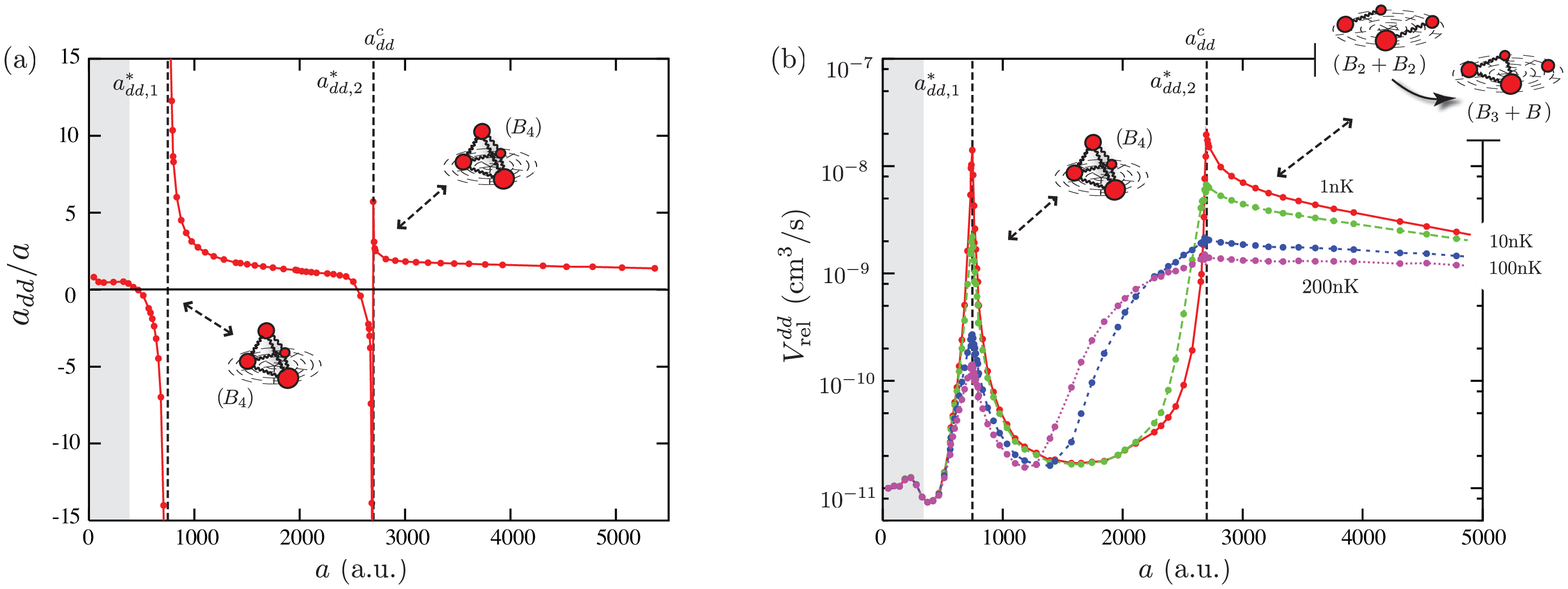}
\caption{(color online). (a) shows poles in the dimer-dimer scattering length $a_{dd}$, plotted as a function of the atom-atom scattering length $a$.  
The poles reflect the emergence of the two universal four-boson
states at $a=a_{dd,1}^*$ and $a_{dd,2}^*$ that are associated with every Efimov trimer[see Fig.~\ref{Fig1}(b)]. These states allow for the control of the dimer-dimer interactions.
(b) shows the thermally-averaged $V_{\rm rel}^{dd}$ and resonant behavior associated with the four-boson states leading to relaxation 
into deeply bound states. For $a>a_{dd}^c$, however, $V_{\rm rel}^{dd}$ is dominated by $B_{2}+B_{2}\rightarrow B_{3}+B$ rearrangement collisions.}
\label{Fig2}
\end{figure*}

A key property resulting from the universality in the four-boson problem is that the aforementioned dimer-dimer resonances
and rearrangement reaction are universally related to the three-body Efimov physics. The values for $a_{dd,i}^*$ and 
$a_{dd}^c$ are controlled by a single three-body parameter, $a_{ad}^*$.
We, therefore, express our results in terms of the universal ratios $a_{dd,i}^*/a_{ad}^*$ and $a_{dd}^c/a_{ad}^*$, which determine the positions of the dimer-dimer resonances, $a_{dd,i}^*$, and the critical scattering length $a_{dd}^c$
at which the $B_{2}+B_{2}\rightarrow B_{3}+B$ rearrangement reaction is enhanced.
In practice, our calculations extract the values for such ratios from the energies in the region near the third Efimov state [see Fig.~\ref{Fig1}(b)]
where the states are largely unaffected by nonuniversal short-range physics. We obtain
\begin{equation}
\frac{a^*_{dd,1}}{a^*_{ad}}\approx2.37,
\mbox{~~~}
\frac{a^*_{dd,2}}{a^*_{ad}}\sim 6.6,
\mbox{~~~and~~~}
\frac{a^c_{dd}}{a^*_{ad}}\approx6.73.
\label{scatratios}
\end{equation}
We expect that such ratios are universal within an error of less than 2\%, as determined
by comparison of our value for $a_{dd}^c/a_{ad}^*$ to the value obtained using the
semi-analytical results from Ref.~\cite{braaten2006ufb}.

In Fig.~\ref{Fig2}, we show our numerical calculations for $a_{dd}$ and for the dimer-dimer relaxation rate, 
$V_{\rm rel}^{dd}$, calculated using: 
\begin{eqnarray}
a_{dd}\stackrel{k\rightarrow0}{=}-\frac{{\rm Re}[\tan\delta_{dd}]}{k}~~{\rm and}~~
V_{\rm rel}^{dd}=\frac{8\pi\hbar}{m k}\sum_{f}|S_{f\leftarrow dd}|^2,\label{ScatOb}
\end{eqnarray}
\noindent
where $k^2=2mE_{col}/\hbar$, with $E_{col}=E-2E_{b}$ being the collision energy, and $\exp({2i\delta_{dd}})=S_{dd,dd}$. The 
$S$-matrix elements are obtained from the solutions of Eq.~({\ref{radeq}}). In our calculations, the only decay channel is the one associated with 
the lowest (relatively deep) Efimov state. We, therefore, do not take into account the decay into deeply bound
two-body states \cite{FFdimers}, but we still expect the general aspects of our results to remain valid.

Figure~\ref{Fig2} shows our numerical results, with the atomic mass chosen to be that of Cs, and $a_{ad}^*=400$~a.u., accordingly with 
the observations of Ref.~\cite{ADRudi}, where $a_{ad}^*=400$~a.u. 
Figure~\ref{Fig2}(a) then shows the dimer-dimer resonances caused by the  four-boson states shown in Fig.~\ref{Fig1}(b). 
We note, however, that from Fig.~\ref{Fig2}(a) we obtain $a_{dd}^c/a_{dd,1}^*\approx3.57$, 
which deviates from our predicted values from Eq.~(\ref{scatratios}),  $a_{dd}^c/a_{dd,1}^*\approx2.84$, in about $20\%$. These deviations are produced by 
finite-range effects that introduce nonuniversal corrections \cite{JPB} to the energies of the two-, three- and four-body systems in Fig.~1(b). 
Figure~\ref{Fig2}(b) shows our results for the thermally averaged $V_{\rm rel}^{dd}$. It clearly exhibits an enhancement of the inelastic loss 
at $a_{dd,1}^*$, corresponding to the dimer-dimer resonance discussed above. Near $a_{dd,2}^*$($\approx a_{dd}^c$), however, the resonance effect 
is masked by an enhanced rate for the rearrangement reaction $B_{2}+B_{2}\rightarrow B_{3}+B$.
In fact, our calculations show an efficiency of about $98\%$ for the
$B_{2}+B_{2}\rightarrow B_{3}+B$ reaction. 
In the presence of deeply bound two-body states, we still expect a high dimer-trimer conversion efficiency
because of the stronger overlap between these states than with other deeply bound channels. 
Therefore, the dominance of the $B_{2}+B_{2}\rightarrow B_{3}+B$ reaction, and its negligible energy released, makes this process 
promising for the formation of Efimov trimers which can (under certain conditions)
remain trapped. In a pure molecular sample, a clear experimental signature of such rearrangement reactions, and therefore of trimer formation, 
would simply be achieved by detecting the reappearance of atoms for $a>a_{dd}^c$.
The lifetime of such trimer states can, however, be a critical issue, since it depends on its
intrinsic decay rate \cite{FXdimers,Felix} and on collisions with other atoms.

Our results can be compared to the experimental data for Cs$_{2}$ of Ref.~\cite{CSdimers} in order to understand the observed scattering length 
and temperature dependences. In Ref.~\cite{CSdimers} it was found that $V^{dd}_{\rm rel}$ exhibits a minimum for relatively large values
of $a$, and that for the largest values of $a$ it was found that $V_{\rm rel}^{dd}$ increases with the temperature, in contrast to the constant
behavior predicted by the Wigner threshold law. Based on the temperature dependence, which we discuss in details below, 
we believe that the minimum observed in Ref~\cite{CSdimers} could be associated with our results in the region $a_{dd,1}^*<a<a_{dd}^c$. 
Even though the range of $a$ in which the minimum is observed in $V^{dd}_{\rm rel}$ differs from that observed experimentally, finite range corrections \cite{JPB} can be expected to be more severe for the experimental data than for our numerical calculations, performed much deeper in the $a\gg r_{0}$ universal 
regime. 

In order to better understand the physics underlying the intriguing temperature dependence found in Ref.~\cite{CSdimers},
we present a simple model for relaxation that captures the important physics also seen in our numerical calculations [Fig.~\ref{Fig2}(b)].
The source of the temperature dependence in our model, and calculations, is related to the presence of a trimer state close to the dimer-dimer 
collision threshold. 
It is well known that the Wigner threshold law is applicable only for energies (temperatures) smaller than the smallest energy scale 
in the system \cite{Limits}. Therefore, any deviation from the Wigner threshold law can 
only be understood in  terms of a new energy scale in the system that is  smaller than the characteristic energy scale given by 
$E_{b}\approx\hbar/ma^2$ \cite{EbCs}. 
Based on the knowledge of the spectrum for Cs$_{2}$ \cite{Mark}, this new energy scale can be associated with a trimer 
state, allowing for the enhancement of the $B_{2}+B_{2}\rightarrow B_{3}+B$ reaction as the temperature 
is increased [Fig.~\ref{Fig2}(b)]. 

\begin{figure}[htbp]
\includegraphics[width=3.2in,angle=0,clip=true]{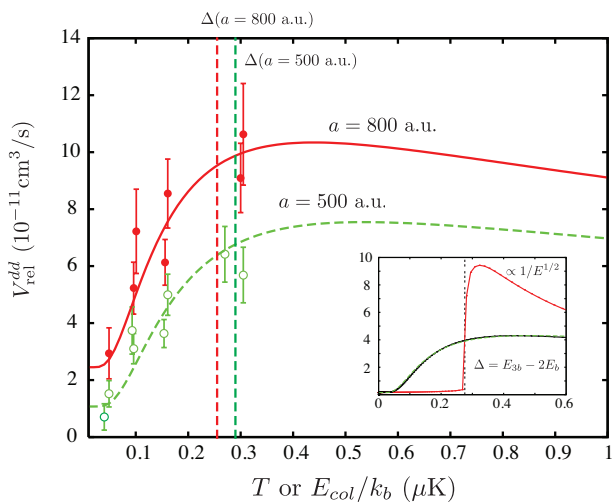}
\caption{(color online). Comparison between experimental data \cite{CSdimers} (filled circles: $a=800$~a.u.; open circle: $a=500$~a.u.) and our model [Eq.~(\ref{VrelT})]
for the temperature dependence of $V_{\rm rel}^{dd}$. 
In our model, the apparent deviation from the Wigner threshold law is caused by the presence of a trimer state just above the dimer-dimer threshold, as indicated by the vertical lines. Inset: Comparison of our model 
(black-solid line) with numerical rates (the red-solid line is our total rate as a function of $E_{col}$, and the green-dashed line shows the thermally averaged 
results as a function of $T$).}
\label{Fig3}
\end{figure}

In our model, therefore, the important energy scale in the system will be $\Delta=|E_{3b}-2E_{b}|<E_{b}$, associated with the energy of the trimer state.
For $E_{col}<\Delta$, our simple model assumes a Wigner threshold law behavior, that $V_{\rm rel}^{dd}=\hbar c_{\eta}/m$, 
where $c_{\eta}$ has dimensions of length. For $E_{col}>\Delta$, however, we assume that the rate is unitarity-limited, 
$V_{\rm rel}^{dd }=f_{\eta}\hbar 4\pi/({2}^{\frac{1}{2}}m^{\frac{3}{2}})E_{col}^{-1/2}$, with $f_{\eta}\leq1$. Its thermal average is: 
\begin{eqnarray}
V_{\rm rel}^{dd}(T) &= &\left[{\rm Erf}\left[\left(\frac{\Delta}{k_{b}T}\right)^{\frac{1}{2}}\right]
-\left(\frac{\Delta}{k_{b}T}\right)^{\frac{1}{2}}\frac{2~e^{-\Delta/k_{b}T}}{{\pi}^{\frac{1}{2}}}\right]\frac{\hbar c_{\eta}}{m}\nonumber\\
&+&\frac{8\hbar\sqrt{2\pi}~e^{-\Delta/k_{b}T}}{m^{3/2}(k_{b}T)^{1/2}}f_{\eta},
\label{VrelT}
\end{eqnarray}
where $k_{b}$ is the Boltzman's constant. The inset of Fig.~\ref{Fig3} shows a comparison between Eq.~(\ref{VrelT})
and our numerical calculations for $a\lesssim a_{dd}^{c}$ and establishes the validity of our analytical model. 
We then use Eq.~(\ref{VrelT}) 
to fit the experimental data shown in Fig.~\ref{Fig3} and find excellent agreement.
However, since our model does not specify the nature of the trimer state, we can not determine if the temperature dependence in Ref.~\cite{CSdimers} is caused by an Efimov state.

In summary, we have demonstrated the universal properties of ultracold dimer-dimer
collisions and determined their connection with three-body Efimov physics. The existence of universal
resonances allows for the tunability of the dimer-dimer interaction 
and opens up the possibility of study molecular gases with attractive and repulsive interactions. 
Our numerical results show that ultracold molecular gases might offer an efficient path for the
creation of a gas of Efimov trimers via $B_{2}+B_{2}\rightarrow B_{3}+B$ rearrangement reactions.
In particular, the observation of resonant effects in dimer-dimer collisions obeying 
the ratios $a^*_{dd,i}/a^*_{ad}$ and $a^c_{dd}/a^*_{ad}$, within the limits imposed by nonuniversal corrections we discussed here, is hereby shown to provide a simultaneous verification of 
both three- and four-body universal physics.

This work was supported in part by the National Science Foundation.
We thank the group of R. Grimm for sharing their experimental data and for
stimulating discussions. The Keck Foundation provided computational resources.


\begin{thebibliography}{99}
\bibitem{EbCs} In fact, for Cs the corresponding values for $E_{b}$ \cite{Mark}
are much higher than the temperatures in Ref.~\cite{CSdimers} and therefore the threshold law should apply.
\end{thebibliography}
\end{document}